\documentclass[runningheads]{llncs}
\usepackage[T1]{fontenc}
\usepackage{graphicx}
\usepackage{amssymb}
\usepackage{amsmath}
\usepackage{color}
\usepackage{caption}
\usepackage{xfrac}
\usepackage{hyperref}
\hypersetup{
    breaklinks=true,
}
\usepackage[anythingbreaks]{breakurl}
\usepackage{todonotes}

\usepackage{multicol, multirow}
\usepackage{booktabs}
\begin{document}
\title{Revealing the empirical flexibility of gas units through deep clustering}
%
%
\author{Chiara Fusar Bassini\inst{1,2}\orcidID{0009-0004-2346-9132} \and
Alice Lixuan Xu\inst{1}\orcidID{0009-0002-4743-4992} \and
Jorge Sánchez Canales\inst{1}\orcidID{0009-0005-7011-3021} \and
Lion Hirth\inst{1} \orcidID{0000-0002-5563-3310}\and 
Lynn H. Kaack\inst{1,2} \orcidID{0000-0003-3630-3102}}
\authorrunning{Fusar Bassini et al.}
%
\institute{Centre for Sustainability, Hertie School, Friedrichstraße 180, 10117, Germany \and
Data Science Lab, Hertie School, Friedrichstraße 180, 10117, Germany
\email{\{c.fusarbassini,l.xu, j.sanchez-canales, hirth, kaack\}@hertie-school.org}}
\maketitle              

\begin{abstract}The flexibility of a power generation unit determines how quickly and often it can ramp up or down. In energy models, it depends on assumptions on the technical characteristics of the unit, such as its installed capacity or turbine technology. In this paper, we learn the empirical flexibility of gas units from their electricity generation, revealing how real-world limitations can lead to substantial differences between units with similar technical characteristics. Using a novel deep clustering approach, we transform 5 years (2019-2023) of unit-level hourly generation data for 49 German units from 100 MWp of installed capacity into low-dimensional embeddings. Our unsupervised approach identifies two clusters of peaker units (high flexibility) and two clusters of non-peaker units (low flexibility). The estimated ramp rates of non-peakers, which constitute half of the sample, display a low empirical flexibility, comparable to coal units. Non-peakers, predominantly owned by industry and municipal utilities, show limited response to low residual load and negative prices, generating on average 1.3 GWh during those hours. As the transition to renewables increases market variability, regulatory changes will be needed to unlock this flexibility potential.
\end{abstract}

\section{Introduction}\label{sec:introduction} 
Ensuring adequate flexibility of conventional power generation units is critical in electricity systems with a high penetration of renewable energy sources (RES). Flexible units that can operate in a high cycling mode, meaning that they can adapt to market conditions by rapidly switching on and off, can better complement the daily fluctuations of RES generation. In energy models, assumptions about the flexibility of a generation unit are often based on technical attributes. For example, the assumed maximum ramp rate for a gas unit will depend on its size, age, and turbine technology (combined or open-cycle): an older combined-cycle plant will be assumed to be less flexible than an open-cycle one. 
\newline \newline
However, the technical capability of a power generation unit (how flexibly it can operate) can starkly differ from its empirical dispatch (how flexibly it operates in practice). Contract obligations, co-generation and economic considerations can significantly limit its empirical flexibility. Hermans and Delarue~\cite{hermans} show that maintenance contracts between plant operators and turbine manufacturers may include penalties for frequent or fast cycling, discouraging flexible generation. Cross-model comparisons~\cite{bucksteeg2022transformation}\cite{ruhnau2022electricity} highlight the sensitivity of energy models to the utilization rate of CHP components, i.e.,~the extent to which a unit is assumed to co-generate heat and electricity. Beykirch et al.~\cite{beykirch2018learning}\cite{beykirch2019bayesian} demonstrate that adjusting the cost parameters of gas-fired generation units based on their dispatch can explain the differences in empirical flexibility between two gas-fired units with similar technical characteristics. 
\newline \newline
To estimate empirical flexibility, we need to complement assumptions based on the technical characteristics of a generation unit with evidence from its observed dispatch, which more closely reflects existing constraints. In this study, we group units of similar empirical flexibility by observing their dispatch over a prolonged time period. We analyze 49 German gas units of at least 100 MWp, responsible for more than 60\% of the yearly national gas-fired generation, whose hourly dispatch is reported on ENTSO-E Transparency~\cite{entsoe_gen}. We consider a five-year period characterized by several market changes (2019 - 2023): a rapid transition to renewables, the shutdown of nuclear and coal plants, and the 2021-2022 gas crisis, which led to exceptionally high gas and electricity prices in many European markets. 
\newline \newline
To process high-dimensional time series, we adapt a neural network to create embeddings of time series of hourly generation, which are easier to compare and cluster. We obtain four groups based on the empirical operation of sample units, and analyze the most relevant techno-economic characteristics of each cluster. Our unsupervised method reveals two types of clusters: fast-ramping, market-oriented units (peakers), and non-peakers units with limited empirical flexibility. We find that the generation of non-peakers, around half of the units, is not significantly correlated with national load, and constitutes over 83\% of sample generation during negative price hours. Our findings underscore that empirical flexibility is not solely determined by technical characteristics and is shaped by market preferences and economic incentives. 
\newline \newline
The paper is structured as follows. In Section~\ref{sec:gas}, we provide a brief overview of the German market. In Section~\ref{sec:methods}, we describe the deep clustering approach used to obtain clusters of gas-fired generation units. In Section~\ref{sec:results}, we examine the resulting clusters and highlight the differences between peaker and non-peaker units. We discuss the policy implications and limitations of our results in Section~\ref{sec:discussion} and conclude in Section~\ref{sec:conclusion}.
\section{Context}\label{sec:gas}
\begin{figure}
    \centering
    \includegraphics[width=.6\linewidth]{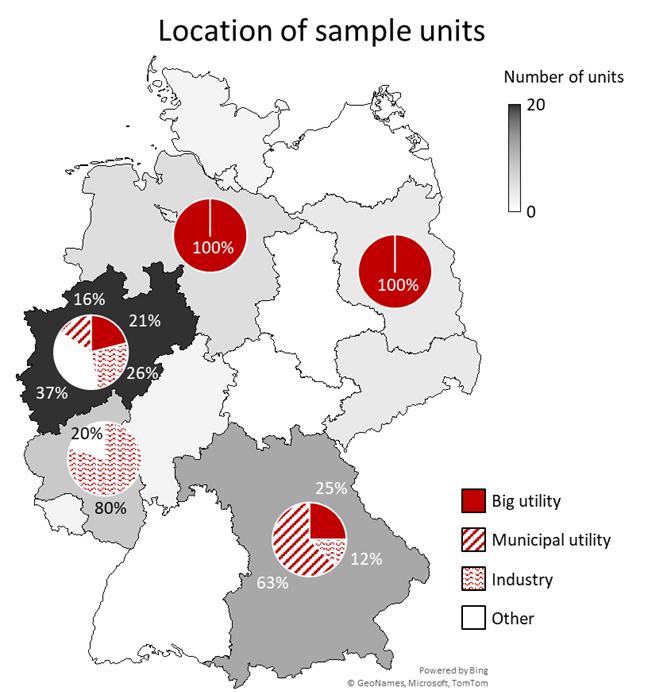}
    \caption{Number of gas units in the sample by state and ownership. Ownership categories are described in Section~\ref{sec:gas}.} 
    \label{fig:map}
\end{figure}

To meet its 2050 carbon neutrality goals, the European Union has adopted an extensive energy transition strategy~\cite{EUCommission2023}. Natural gas is considered a "bridge fuel" that can facilitate the phase-out of coal-fired units and balance RES variability. Therefore, many EU countries heavily invested in gas generation and distribution networks; the German government, in particular, has advocated for expanding its fleet of gas-fired generation units to support the strong growth in variable generation~\cite{bmwk}. Less attention was paid to the evaluation of existing units, whose total capacity in 2024 amounted to 36.7 MWp, or 48.6\% of the capacity of all conventional assets~\cite{energy_charts}.
\newline \newline
Between 2015 and 2021, the share of natural gas in the German electricity generation mix increased from 2.9\% to 10.1\%, and gas-fired units went from being rarely price-setting (less than 3\%) to setting prices nearly 24\% of the time~\cite{ZAKERI20232778}. In particular, during 2019 and 2020, higher coal costs accelerated the switch to gas-fired generation, which became a cheaper alternative to coal-based generation in many west European countries~\cite{BloombergNEF2023}. However, between 2021 and 2022, Russia significantly reduced its natural gas exports to Europe, triggering a severe energy crisis, which affected the German economy and the electricity sector~\cite{ruhnau2023gas}. These market changes are reshaping the way gas-fired generation units need to be dispatched.
\newline \newline
Generation companies often react differently to market changes~\cite{barazza}. To identify the owners of German gas units, we obtain new data from EEX~\cite{eex_list}. Owners are then classified into major utilities, municipal utilities, industrial companies, and other utilities, where the latter includes regional and foreign companies of medium dimension. Since generation units can have multiple owners, we assume the company reporting on EEX to be its main stakeholder, using 2021 as the reference year. The five major utilities (RWE, EnBW, Uniper, Vattenfall and LEAG) accounted for 66.9\% of the national unsubsidized electricity generation in 2021~\cite{Marktmachtbericht}. Municipal utilities are small local entities (around 1,000 entities~\cite{stadtwerke}) responsible for local energy and heat production, water and waste management. Energy-intensive industries also own power generation units, co-located within their production sites. As shown in Figure~\ref{fig:map}, most industrial units are located in the South West of the country.
\section{Methodology}\label{sec:methods} 
To cluster the hourly operation of multiple power generation units over several years, we need to compare many high-dimensional time series. Traditional clustering methods struggle in such a setting: Euclidean distance is sensitive to temporal shifts, while dynamic time warping suffers from poor scalability~\cite{rani2012recent}\cite{dtw}. To address these limitations, we resort to deep clustering, a class of methods that cluster high-dimensional data with deep neural networks~~\cite{zhou2024comprehensive}. 
\newline \newline
Deep clustering methods entail a representation learning module, which leverages a deep neural network to compress raw data to low-dimensional representations, and a clustering module, which clusters these representations. Previous energy-related research has mainly deployed deep clustering for residential load profiling and RES forecasting~\cite{wang2018deep}\cite{eskandarnia2022embedded}\cite{han2020short}, while other applications remain largely unexplored. 
\begin{figure}[t]
     \includegraphics[width=1\linewidth]{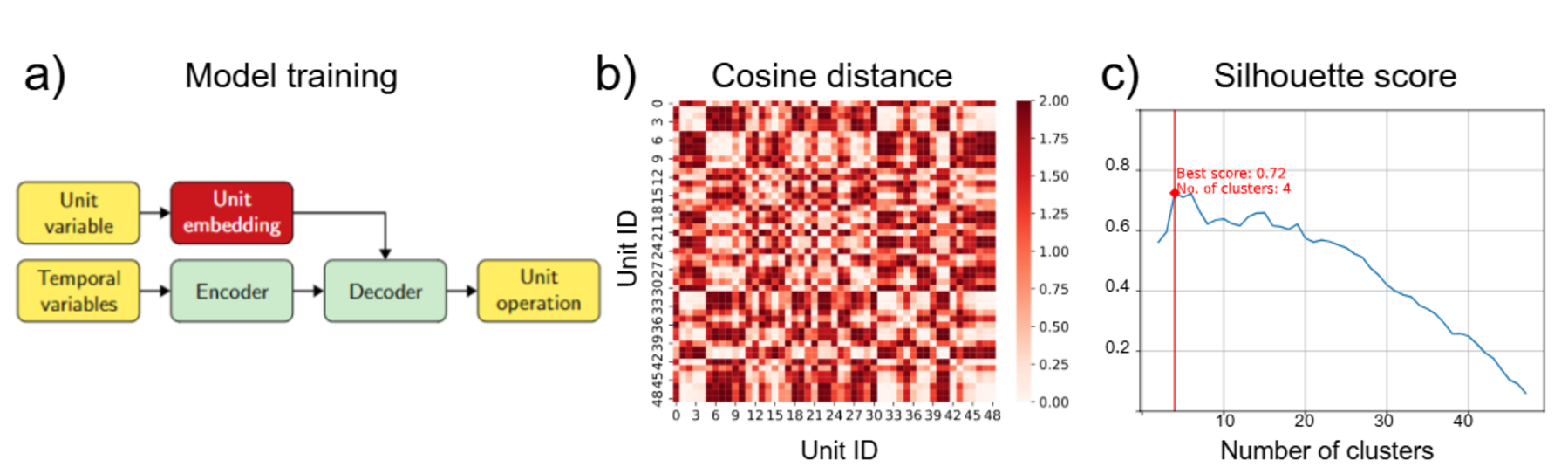}
    \caption{Model pipeline: a) train a deep neural network to predict unit operation, b) cluster the model embeddings using their cosine distance matrix, and c) select optimal cluster number with the Silhouette Score.}
\label{fig:model_pipeline}
\end{figure}
\newline \newline
Building on an encoder-decoder neural network from Mirzavand et al.~\cite{mirzavand2023explainable}, we design a deep clustering procedure which is fully unsupervised and does not make use of the techno-economic characteristics of the units. The clustering pipeline entails the following steps: first, we train the neural network to predict the next-day operation of the gas-fired units (Figure~\ref{fig:model_pipeline} a), obtaining a test accuracy comparable to that of the original model from Mirzavand et al.~(84.6\%). We construct a matrix of distances from the embeddings of the trained model, which is used for hierarchical clustering (Figure~\ref{fig:model_pipeline} b). The number of clusters is selected through Silhouette Score maximization, balancing the cohesion and separation between clusters; four is identified as the optimal number of clusters (Figure~\ref{fig:model_pipeline} c).
\subsection{Data processing}
Our primary data are disaggregated hourly generation of 49 gas-fired units with an installed capacity of 100 MWp or more, published on ENTSO-E Transparency~\cite{entsoe_gen}. We complement these data with the total national load and RES generation~\cite{entsoe_dem}\cite{entsoe_res}, gas prices~\cite{dutch_ttf}, as well as one-hot encoded time features (weekday, hour). We use five years of data (2019–2023), resampled to hourly granularity and scaled using min-max normalization. Following Beykirch et al.~\cite{beykirch2018learning}\cite{beykirch2019bayesian}, we discretize generation into states:
$$
y_{it} = 
    \begin{cases}
    0 & g_{it} \leq \frac{1}{3} \; G^{\max}_i \\
    1 & \frac{1}{3} \; G^{\max}_i < g_{it} \leq \frac{2}{3} \; G^{\max}_i \\
    2 & g_{it} > \frac{2}{3} \; G^{\max}_i, \\
    \end{cases}
$$
where $g_{it}$ is the generation of unit $i$ at time $t$, $G^{\max}_i$ its installed capacity and $y_{it}$ the model input. We exclude units that were almost never in operation, as we assume them to be used only as capacity reserves. To identify them, we evaluate the unit-level entropy of the input distribution:
$$
H_i = - \sum_{k=0}^2  p(y_{it} = k) \log_2 p(y_{it} = k),
$$
where $p(y_{it} = k)$ is the relative frequency of state $k$ (0: off, 1: min-load, 2: full-load) for unit $i$. We select a sample of 49 units with a minimum entropy of 0.3: units with a lower entropy are off for more than 95\% of the five-year period.
\newline \newline
We process the input data into temporal subsequences of 48 hours of unit observation and a one-hot encoded unit identifier. The context window (24 hours) includes both the previous states for the unit and the min-max scaled temporal variables (encoder input), while the forecasting window of 24 hours contains temporal variables only (decoder input). The context length was set through hyperparameter tuning, while the forecast length aligns with European regulation\footnote{Power plant operators are required submit generation forecasts for the next 24h by the Transmission System Operator~\cite{sys_guideline}.}. We use a step size of 48 hours between subsequences, so that they do not overlap. The final dataset contains 42,913 subsequences, which are randomly split into training, validation, and testing with a ratio of 70/20/10. 
\subsection{Representation module}
Building upon the model architecture from Mirzavand et al.~\cite{mirzavand2023explainable}, we design an encoder-decoder model that forecasts the 24 day-ahead operation of a generation unit, receiving as input its historical operation and aggregated market variables. The encoder and decoder layers contain two Gated Recurrent Units (GRU). At each iteration step, the model predicts the state of one generation unit (0: off, 1: min-load, 2: full-load) for the next 24 hours, based on past states and temporal variables for the previous 24 hours. Our deep clustering approach extends the decoder module, adding an embedding layer that captures unit-specific patterns in a compressed form.
\newline \newline 
The encoder block processes the temporal input into a fixed-dimensional hidden state that is fed to the decoder. The decoder block also receives a static one-hot encoded vector, which serves as the unit identifier, so that it can distinguish which unit the input time series belongs to. Instead of directly using the one-hot encodings, we transform them to lower dimensional vectors using a matrix learnable embeddings. Embeddings are initialized as zero vectors, learned by the model during training, and then used as input for deep clustering. The embedding matrix has dimension $\mathbb{R}^{m\times n}$, where $n$ is the number of units and $m$ the length of the embedding vectors, which is set to 8 after hyperparameter tuning. Ultimately, the decoder block generates its predictions step by step, and these are passed through an output layer to predict the output classes. 
\newline \newline
The model is trained by minimizing the cross-entropy loss using the Adam optimizer~\cite{diederik2014adam}. Given the skewness of the class distributions, we monitor both regular and balanced accuracy, i.e.,~the average recall over all classes. The final model achieves a similar forecasting performance to the model from Mirzawand et al.~\cite{mirzavand2023explainable} with one-hot encodings, with a test accuracy of 84.6\% (original model: 84.4\%) and a balanced accuracy of 78.0\% (original model: 78.4\%). Its hyperparameter configuration is reported in the Appendix Table~\ref{sec:parameters}. 
\subsection{Clustering module}
We use the unit embeddings obtained from the model training to derive clusters. To this end, we apply hierarchical clustering, which has already been successfully deployed for deep clustering, e.g.,~in Yang et al.~\cite{yang2016joint}. At initialization, each cluster contains a single unit; the algorithm iteratively merges the clusters that minimize the farthest distance between any pair of points (complete linkage). We use complete linkage to mitigate the chaining phenomenon and obtain more even-sized clusters~\cite{yim2015hierarchical}. \newline \newline 
Hierarchical clustering can be paired to different (not necessarily Minkowski) distance metrics, and matrices of pairwise distances between units. Following previous literature on embedding distances~\cite{cassisi2012similarity}\cite{bolukbasi2016man}, we use the cosine distance, i.e.,~the complement of cosine similarity:
\begin{equation*}
  d_{ij}= 1 - \frac{e_i e_j}{\|e_i\|\|e_j\|}, \label{eq:cosine}
\end{equation*}
where $e_i$ is the embedding vector of unit $i$, and $n$ the number of units in the sample. We build a distance matrix $D \in \mathbb{R}^{n \times n}$ (Figure~\ref{fig:model_pipeline} b), with entries between 0 and 2 (included). If $d_{ij}= 1$, embeddings are orthogonal, while 0 and 2 correspond, respectively, to exactly identical and opposite vectors.
\subsection{Robustness checks}\label{subsec:robustness}
We assessed the robustness of our deep clustering approach by varying either the observation period used as input or the clustering algorithm, while keeping the number of clusters fixed to 4. First, as we expected that the gas crisis had impacted the operation of gas-fired units, we divided the dataset into two time periods, pre-crisis (01/2019 - 06/2021) and crisis (07/2021 - 12/2023), and derived new clusters using these temporal subsets. Second, to assess the impact of varying the clustering module, we compared results obtained with K-Means and hierarchical clustering with single or average linkage. We measured the consistency of the cluster assignments between the experiments by the level of cluster agreement, i.e.,~the share of units that are in the same cluster in both configurations.
\newline \newline
Finally, we tested the effectiveness of other clustering methods on the data. In particular, we tried two raw-data-based methods, which cluster time series in a single-step procedure, and one feature-based method, which transform time series to another domain -- e.g.,~frequency~\cite{agrawal1993efficient}. For the raw-based category, we used $K$-Means in combination with Dynamic Time Warping (DTW), which computes the optimal alignment between two sequences~\cite{dtw}, and a single-step, correlation-based hierarchical clustering. Since correlation is not a distance measure, we applied linear transformation $d(x,y)=1-\text{corr}(x,y)$ instead. For the feature-based category, we mapped the time series of scaled generation to their first $n$ Fourier coefficients and clustered them with $K$-Means. 

\section{Results}\label{sec:results} 
\begin{figure}[htpb]
    \centering
    \includegraphics[width=\linewidth]{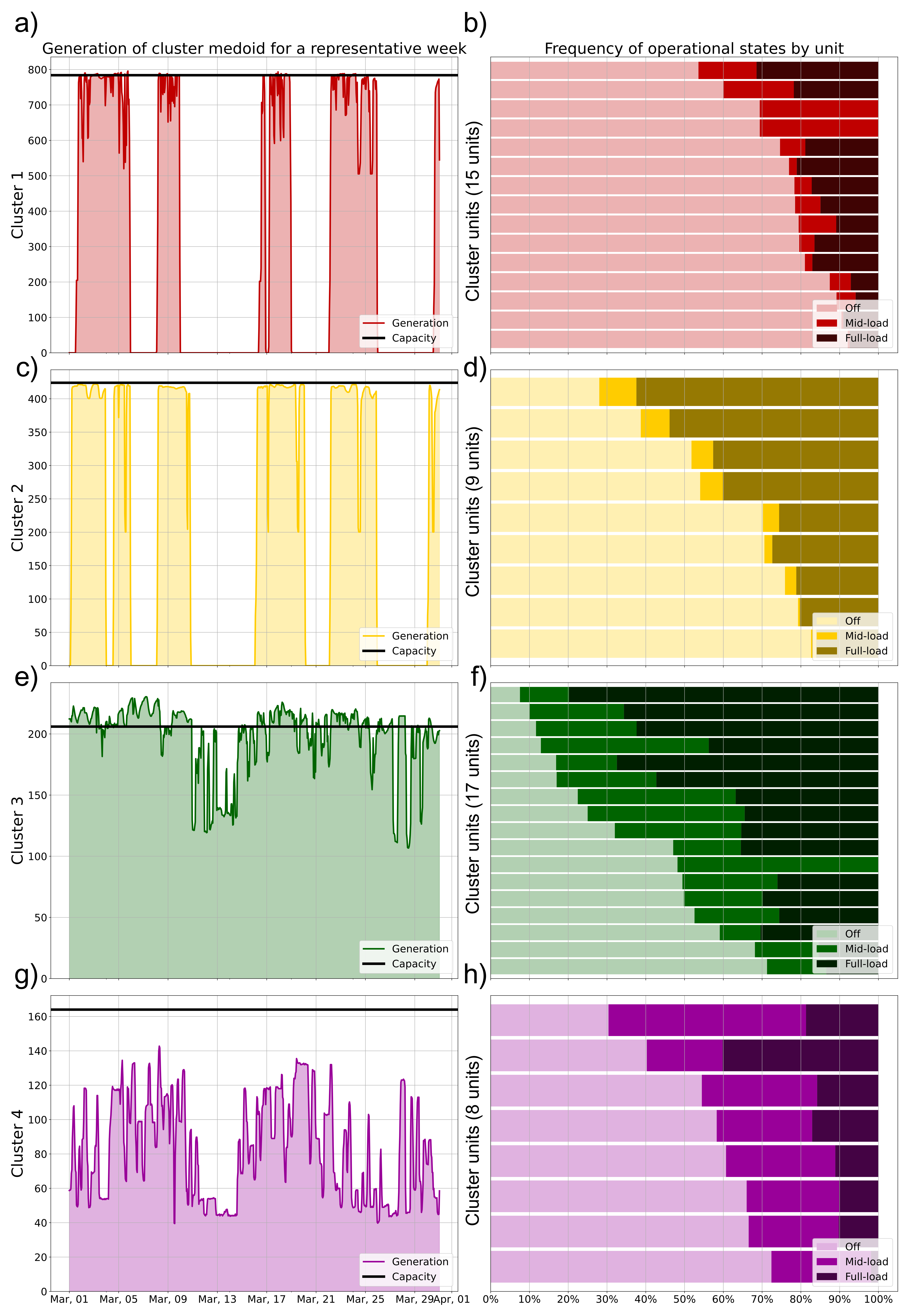}
    \caption{Left: generation of cluster medoids for a representative week. Right: frequency of states (off, min-load, full-load) over total hours of each unit, by cluster.}
    \label{fig:what_model_sees}
\end{figure}

Our unsupervised method reveals four clusters of gas-fired generation units in Germany, displayed in Figure~\ref{fig:what_model_sees}. Clusters 1 and 2, representing 49.0\% of the units, include peaker units with high empirical flexibility. The remaining units in Clusters 3 and 4 are non-peakers, are less flexible and maintain relatively stable dispatch levels over time. 
\subsection{Cluster composition}\label{subsec:cluster_statistics}
The clusters, derived by grouping units based on their dispatch, exhibit different levels of empirical flexibility, as visible in Figure~\ref{fig:what_model_sees}. We use a hand-curated dataset, which includes two public sources~\cite{mastr,opsd} and novel data from EEX~\cite{eex_list}, to analyze cluster composition. In particular, we examine the technical characteristics of the clustered units, such as age, capacity (MWp), turbine technology and economic characteristics, such as ownership and market orientation, i.e.,~whether they are active on the day-ahead market. Cluster-level averages of unit characteristics are reported in Table~\ref{sec:cluster_stats}.
\newline \newline
Cluster 1 includes 15 units, which are mainly owned by major and mid-sized utilities (80.0\%); none of the units is owned by an industrial company. They are all market-oriented (their output is traded on the EEX platform) and relatively modern: 73.3\% of them are 20 years or less. This cluster contains occasional peakers, with an average share of full-load hours of only 13.3\%. Cluster 1 has the highest average market values, which exceed yearly sample averages by 10–12\%. Between 2019 and 2023, half of the top 10 sample units by market value belonged to this cluster. This confirms their peaker role, i.e.,~they mostly generate when residual load is high, and therefore obtain higher revenues. 
\newline \newline
Cluster 2 displays similar characteristics to Cluster 1. Its 9 units are mainly owned by major or mid-sized utilities: although there is one industrial unit in the cluster, it is managed by a major utility~\cite{rwe_ludwigshafen}. Similarly to Cluster 1, units in Cluster 2 are relatively younger and larger, with a mean installed capacity of 396.1 MWp (+33.7\% compared to the sample mean). However, units in Cluster 2 generate more frequently, with an average share of hours at full-load of 34.4\% (Cluster 1: 13.3\%). Cluster-level yearly market values are consistently above sample-level averages, yet slightly below market values for Cluster 1 (8\%-11\% lower for all years except 2020).
\newline \newline
Cluster 3 contains 17 non-peaker units and is the largest by number of units. 52.9\% of the units in Cluster 3 are owned by municipal utilities. This includes units in the cities of Berlin and Hamburg, owned between 2019 and 2023 by the municipal subsidiary of Vattenfall, a big utility~\cite{vattenfall2024}. Units in this cluster are likely to be smaller: the majority of them have a capacity of below 250 MWp. Most of the units in Cluster 3 are also equipped with a CHP component (88.2\%). Their generation is not strongly correlated with residual load (0.29 on average), and their higher correlation with heat demand (0.64 on average) is consistent with district heating obligations. Since no nation-wide heat demand data are available for Germany, we use synthetic data from Ruhnau et al.~\cite{ruhnau2019time}.
\newline \newline
Cluster 4 contains predominantly industrial units. As shown in Figure~\ref{fig:what_model_sees} h, these 8 units are mostly off (56.1\% of the time) or dispatched at min-load (28.2\%). Cluster 4 also shows the lowest average correlation with wholesale electricity prices (0.11). Interestingly, our method captures a distinction between industrial facilities of the chemical and refinery industry (mainly in Cluster 3) and those of the steel industry (mainly in Cluster 4). Anecdotal evidence from power plant owners reveals that these units mostly use locally produced coal-based gas, a by-product of iron-related industrial processes, to supply electricity to the co-located facilities.
\newline \newline
Finally, we apply ANOVA and Chi-square tests to determine which technical and economic variables are more distinct between clusters. These statistical tests determine whether observed differences in the distribution of a variable between groups are statistically significant to a predefined $p$-value or negligible. Particularly distinct between clusters are the age of the gas units ($p$-value $ \leq 0.001$) and their market orientation ($p$-value $ \leq 0.01$). The type of owner (industry, major utility, municipal utility, or other utility), capacity, and the presence of a CHP component have a lower significance ($p$-value $ \leq 0.01$), while the technology type (combined-cycle, steam or gas turbine) is not statistically significant.
\subsection{Estimating empirical flexibility}\label{subsec:flexibility}
\begin{figure}[htbp]
    \centering
    \includegraphics[width=\linewidth]{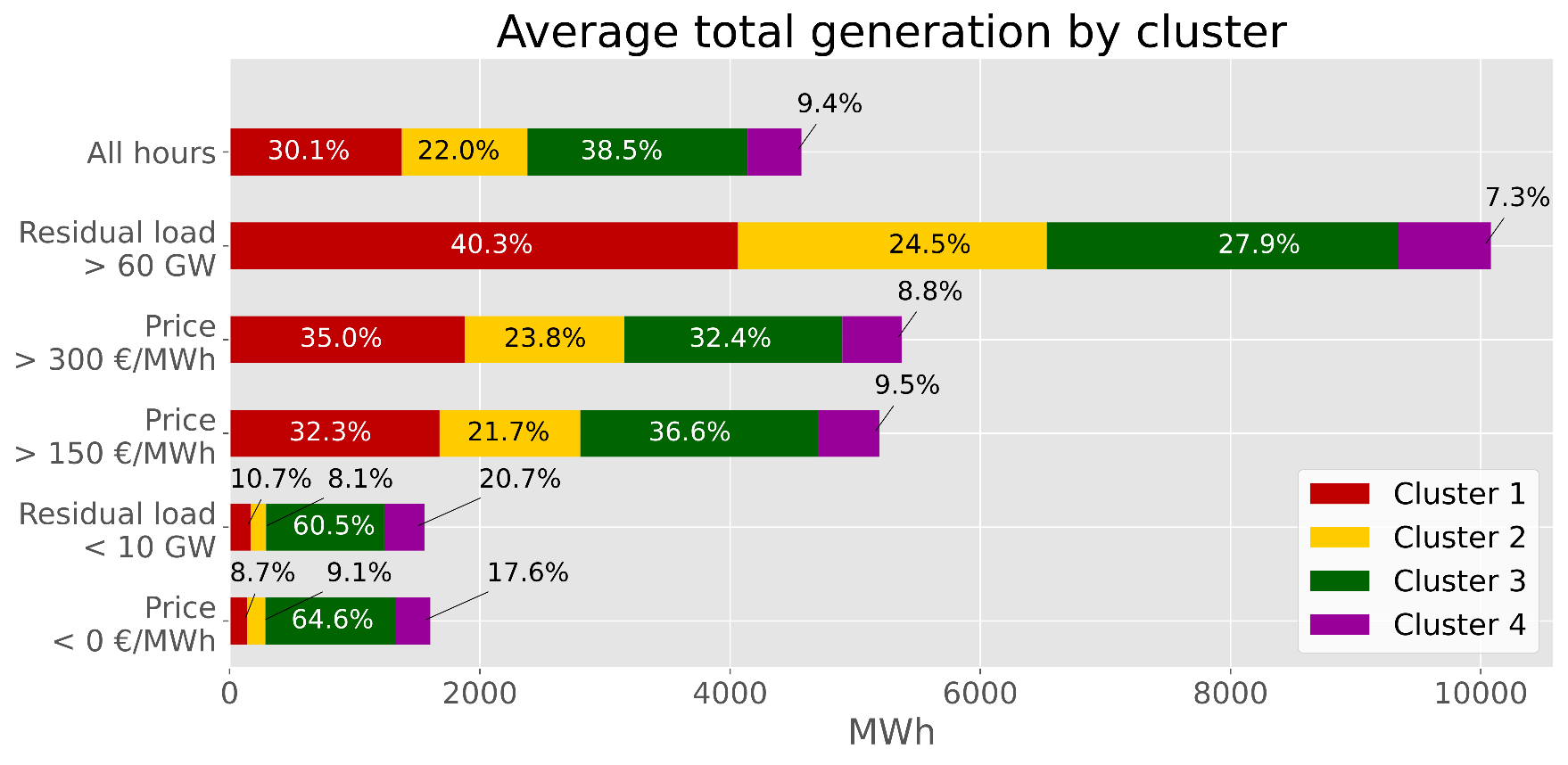}
    \caption{Total generation of sample units and relative cluster shares under different market conditions. Peaker share increases when prices or residual load are higher, decreases with low load and negative prices.}
    \label{fig:avg_generation}
\end{figure}

For each unit, we compute the empirical maximum ramp rate as the maximum observed change in output (MWh) per hour in 2019-2023. In Table~\ref{tab:max_ramp_rates}, we report cluster-level averages and compare the averages with model assumptions for coal and gas units from Krishnamurthy et al.~\cite{Krishnamurthy}. While the ramp rates for Cluster 1 and 2 are in line with the assumed rates for gas units, those for Cluster 3 and 4 are closer to values for coal units. This suggests that the empirical flexibility of non-peakers units may be overestimated in current models.
\newline \newline
\begin{minipage}[t]{0.55\linewidth}
\vspace{0pt}
As RES penetration grows, it becomes more and more critical to operate conventional units in a flexible way. When demand is low and RES generation is high, conventional must-run generation can contribute to negative prices. Between 2019 and 2023, most of the must-run generation in the sample (83.0\%) was due to non-peaker units, which generated on average 1.3 GWh during low load and negative price hours (Figure~\ref{fig:avg_generation}). This could be in part due to CHP must-run constraints, which previous research estimated to be between 2 and 8 GWh in total~\cite{agora2014}. 
\end{minipage}
\hfill
\begin{minipage}[t]{0.4\linewidth}
\vspace{0pt}
    \centering
    \begin{tabular}{p{2.25cm}p{2.25cm}} \toprule
    \textbf{Fuel type} & \textbf{Ramp rate \newline (MW/min)} \\\midrule
        Cluster 1 & 5.7\\
        Cluster 2 & 6.0\\
         Cluster 3 & 3.2\\
         Cluster 4 & 2.6\\\cmidrule{1-2}
         Natural gas$^\ast$ & 6.7 \\
         Coal$^\ast$ &  2.0 \\
         \bottomrule
    \end{tabular}
    \captionof{table}{Max ramp rates (cluster average). Starred values are from Krishnamurthy et al.~\cite{Krishnamurthy}.}
    \label{tab:max_ramp_rates}
\end{minipage}
\subsection{Clusters remain stable during gas crisis}\label{subsec:sensitivity}
Given the market changes experienced by gas-fired units during the gas crisis, we were interested in whether that would affect cluster assignments. Therefore, as described in~\ref{subsec:robustness}, we built clusters based either only on the pre-crisis (01/2019 - 06/2021) or only on the crisis (07/2021 - 12/2023) periods. Most units remained either peakers or non-peakers during the gas crisis: only 12.2\% of the units modified their dispatch enough to change between a peaker and a non-peaker cluster. However, 22.4\% of the units moved between Cluster 1 and 2, both peaker clusters, if they were generating less during one of the periods. For example, München Süd 1 (turbine no.~2) was not active in 2022 due to modernization work~\cite{sw_muenchen}, and moved from Cluster 2 to Cluster 1. 
\newline \newline
Moreover, peakers responded more strongly to the gas crisis than non-peakers. Their overall generation fell sharply during the crisis (07/2021-12/2023), with respect to pre-crisis period 01/2019 - 06/2021. In particular, Cluster 1 and 2 both reduced their total generation  
by respectively 34.9\% and 34.5\% during the crisis. The decline was less pronounced for Cluster 3 (-14.6\%, with respect to a total generation of 41.6 TWh in the pre-crisis period), whereas Cluster 4 slightly increased its generation, which remained between 3-4 TWh per year. These changes reflect differences in the business models between peaker and non-peaker units. As units in Cluster 3 are likely to predominantly serve the heat demand, and those in Cluster 4 to locally source their fuel, they are relying less on electricity and gas prices.
\newline \newline
Finally, the results are not strongly affected by the choice of clustering algorithm, indicating that the deep neural network successfully encodes the variations between generation units in a lower dimension. In fact, we find that different clustering techniques (hierarchical clustering with average and simple linkage, K-Means with Euclidean and cosine distance) obtain similar clusters to the original configuration, with an agreement level of at least 85.9\%. In line with previous literature on time series clustering~\cite{aghabozorgi2015time,liao2005clustering}, our experiments confirm the advantages of model-based clustering for high-dimensional time series. Raw-based methods performed poorly due to computational cost (DTW distance) or extremely imbalanced clustering (correlation-based). The results of the feature-based method were not robust to different number of Fourier coefficients $n$, possibly because these coefficients are heterogeneous in magnitude and direction. 
\section{Discussion}\label{sec:discussion}
We show that the empirical flexibility of part of the German gas fleet is significantly limited. 51.0\% of the sample operates as non-peaker, with low sensitivity to wholesale electricity prices. New data sourced from EEX confirm our empirical findings, as more than half of non-peaker units (56. 0\%) are predominantly inactive on the spot market (i.e.,~not market-oriented). These considerations should be included in energy models to capture the limited flexibility of certain units, for example through owner-specific optimization strategies~\cite{barazza}. Despite major market changes, clusters are rather stable over time, hence modelers can use the estimated ramp rates to calibrate their assumptions (Table~\ref{subsec:flexibility}).
\newline \newline 
Our results highlight that the empirical flexibility of generation units is also influenced by their ownership. Most non-peaker units are owned by industries, by municipal utilities, or, in Hamburg and Berlin, by big utilities acting as providers of municipal services. Given their obligation to provide heating to local customers, municipal utilities may be unable or less willing to generate flexibly. Industries benefit from consumption-specific grid fee benefits, which were found to limit their overall flexibility~\cite{bockhacker2024stolperstein}. Regulatory adjustments, in particular with respect to grid fees and CHP regulations~\cite{stromnev,kwk}, may be necessary to unlock this flexibility potential. 
\newline \newline 
Some of the limitations of this study are related to the scope of the data. Because only generation units with a minimum installed capacity of 100 MWp are reported on ENTSO-E, our results are not representative for smaller gas units. However, given the low average capacity of these units (13 kWp), such units are likely backup generators for critical infrastructures; thus we expect them to show no price-driven flexibility. Moreover, data on unit-level outages, which are not available in a compatible format, were omitted. It is also not possible to distinguish the end usage of the generated electricity (e.g.,~balancing, local consumption). We refer to Hirth et al.~\cite{hirth2018entso} for a complete review of ENTSO-E Transparency data.
\newline \newline 
Finally, the developed method is a multistage deep clustering procedure, where the representation and the clustering module are optimized separately in a sequential way. Although multistage deep clustering is an intuitive procedure with a long history of applications for temporal data~\cite{zhou2024comprehensive}, its sequential nature might result in representations that are not optimal for the clustering task. End-to-end approaches, such as simultaneous or iterative deep clustering, may be used to improve the procedure. Furthermore, using explainable techniques may improve our understanding of the clustered representations. 
\section{Conclusion}\label{sec:conclusion}
Energy models assume that gas generation units that are technically similar, such as two combined-cycle turbines of similar size, will operate with similar flexibility. In this study, we demonstrate that their empirical flexibility can be substantially different. We present a novel deep clustering method, where an encoder-decoder learns low-dimensional embeddings from the electricity generation of gas units that are used to cluster them in groups with similar empirical flexibility. 
\newline \newline
We cluster 49 German gas-fired units, responsible for more than 60\% of national gas generation, using an unsupervised deep clustering approach. Units are clustered in two groups of peakers (Clusters 1 and 2) and two of non-peakers (Clusters 3 and 4). Peaker units are flexible and price-sensitive, whereas the empirical flexibility of non-peakers appears to be constrained by service obligations, e.g.~district heating (Cluster 3) or industrial generation (Cluster 4). Non-peaker units, often owned by industry and municipal utilities, may also lack incentives to operate flexibly under existing German regulation. 
\newline \newline
Despite exceptionally high gas prices in 2021-2022, part of the gas fleet is relatively insensitive to changes in the wholesale electricity market. Without clustering, this behavior remains hidden and can lead to overconfident conclusions about the market responsiveness of German gas generation. Our method offers a new way to track changes in the role of dispatchable generation over time; future research could apply it to other national contexts, particularly in countries with high shares of RES and CHP generation.
\subsection*{Data availability} 
All original data sources are publicly available. The processed dataset is published under the CC BY 4.0 license on \hyperlink{https://doi.org/10.5281/zenodo.14583598}{Zenodo}.
\subsection*{Code availability} 
The code is published under the GNU General Public License v3.0 license on \hyperlink{https://github.com/chiara-fb/embed-power-plant}{GitHub}.
\subsection*{Author contributions: CRediT}
C.F.B.: formal analysis, conceptualization, methodology, data curation, visualization, writing - original draft, writing - review \& editing, A.L.X.: conceptualization, writing - review \& editing, J.S.C.: conceptualization, writing - review \& editing, L.H.: supervision, funding acquisition, writing - review \& editing L.H.K.: conceptualization, methodology, supervision, funding acquisition, writing - review \& editing.
\subsection*{Acknowledgments} The project was funded by the Federal Ministry of Research, Technology and Space (BMFTR, project code: 16DKWN102) and is part of the German Recovery and Resilience Plan (DARP), financed by NextGenerationEU, the European Union's Recovery and Resilience Facility (ARF). We are grateful for the feedback from Jacqueline Adelowo, Lukas Franken, Shiva Madadkhani, Kim Miskiw, Oliver Ruhnau, Raffaele Sgarlato, Philipp Staudt, and the participants of the Doctoral Workshop of the 13th DACH+ Conference on Energy Informatics.
\subsection*{Disclosure of Interests}
The authors declare no competing interests.

\bibliographystyle{elsarticle-num} 

\bibliography{main}

\newpage
\appendix
\section{Model parameters}\label{sec:parameters}
\begin{table}[htbp]
    \centering
\begin{tabular}{p{2.75cm}p{1.5cm}}
\toprule
\textbf{Hyperparameter} & \textbf{Selected value} \\\midrule
 RNN layers \newline (encoder/decoder) 
& 2 each \\
 Context size \newline (encoder) 
 & 1 day \\
 Hidden layer size 
 & $2^9$ \\
 Embedding size & $2^3$ \\\bottomrule
\end{tabular}
\end{table}
\newpage

\section{Cluster-level statistics}\label{sec:cluster_stats}
{\small
\begin{tabular}{p{3.1cm}*{5}{p{1.6cm}}}
\toprule 
 & \multicolumn{2}{c}{\textbf{Peaker clusters}} & \multicolumn{2}{c}{\textbf{Non-peaker clusters}}    & \textbf{Total} \\\cmidrule(lr){2-3}\cmidrule(lr){4-5}
& (1) & (2) & (3) & (4) & \\\midrule    
\textbf{\textit{Size}} & 15 & 9 & 17 & 8 & 49  \\
\textbf{\textit{CHP}} & 53.3\% &  55.6\% & 88.2\% & 100.0\% & 73.5\% \\
\textbf{\textit{Coal-based gas}} & 0.0\% & 0.0\% & 5.9\% & 50.0\% & 7.8\% \\
\textbf{\textit{Market-oriented}} & 100.0\% & 88.9\% & 47.1\% & 37.5\% & 69.4\%  \\
\textbf{\textit{Age (years)}} & 22.7 & 21.6 & 21.6 & 44.0 & 25.6  \\
\textbf{\textit{Capacity (MWp)}} & 365.6 & 396.1 & 213.0 &  230.5 & 296.2 \\
\midrule 
\textbf{\textit{Market value \newline (€/MWh)}} &&&&&\\
\textit{2019} & 44.7 & 44.3 & 37.5 & 31.5 & 40.0  \\
\textit{2020} & 40.8 & 37.8 & 33.5 & 35.0 & 36.7 \\
\textit{2021} & 123.3 & 121.0 & 103.2 & 88.4 & 110.2  \\
\textit{2022} & 305.3 & 298.0 & 249.3 & 244.4 & 274.6  \\
\textit{2023} & 131.1 & 129.3 & 109.9 & 107.4 & 119.6\\
\textbf{\textit{Market value \newline factor}} &&&&& \\
\textit{2019} & 1.19 & 1.18 & 1.00 & 0.84 & 1.06 \\
\textit{2020} & 1.34 & 1.24 & 1.10 & 1.15 & 1.21 \\
\textit{2021} & 1.27 & 1.25 & 1.07 & 0.91 & 1.14  \\
\textit{2022} & 1.30 & 1.27 & 1.06 & 1.04 & 1.17  \\
\textit{2023} & 1.38 & 1.36 & 1.15 & 1.13 & 1.26  \\
\textbf{\textit{Residual load \newline correlation}} &&&&& \\
\textit{2019} & 0.31 & 0.43 & 0.19 & 0.21 & 0.27  \\
\textit{2020} & 0.44 & 0.45 & 0.26 & 0.21 & 0.34  \\
\textit{2021} & 0.40 & 0.47 & 0.30 & 0.19 & 0.34 \\
\textit{2022} & 0.38 & 0.45 & 0.34 & 0.21 & 0.35 \\
\textit{2023} & 0.32 & 0.34 & 0.34 & 0.20 & 0.31 \\
\textbf{\textit{Heat demand \newline correlation}} &&&&& \\
\textit{2019} & 0.00 & 0.06 & 0.27 & 0.50 & 0.17   \\
\textit{2020} & 0.09 & -0.07 & 0.33 & 0.31 & 0.18\\
\textit{2021} & 0.33 & 0.29 & 0.42 & 0.39 & 0.36\\
\textit{2022} & 0.29 & 0.19 & 0.36 & 0.25 & 0.29  \\
\bottomrule
\end{tabular}
}

\end{document}